\documentclass[twocolumn,preprintn/Users/antiasanchezbotana/Box Sync/draft_scn_mgo/PRL_SUBMISSION/ScN_prb_submission_v7.texumbers,floatfix,showpacs,
prb,superscriptaddress]{revtex4-1}
\usepackage{graphicx}
\usepackage{float}
\usepackage{dcolumn} 
\usepackage{bm}
\usepackage{epsfig}
\usepackage{natbib}
\usepackage{amsmath}
\usepackage{mathtools}
\usepackage{listings}
\usepackage{longtable}
\usepackage{csquotes}
\usepackage{gensymb}
\usepackage{xcolor}
\usepackage{hyperref}
\usepackage{comment}
\hypersetup{
     colorlinks   = true,
     citecolor    = blue
}

\newcommand{\out}[1]{}

\begin{document} 

\title{All $3d$ electron-hole bilayers in CrN/MgO(111) multilayers}

\author{Antia S. Botana}
\address{Department of Physics, University of California Davis,  Davis, California 95616, USA}
\author{Victor Pardo}
\affiliation{Departamento de Fisica Aplicada, Universidade de Santiago de Compostela, E-15782 Santiago de Compostela, Spain}
\affiliation{Instituto de Investigacions Tecnoloxicas, Universidade de Santiago de Compostela, E-15782 Santiago de Compostela, Spain}
\author{Warren E. Pickett}
\address{Department of Physics, University of California Davis,  Davis, California 95616, USA}

\begin{abstract}

CrN/MgO(111) multilayers modeled via \textit{ab initio} calculations give rise to nanoscale, scalable, spatially separated two-dimensional electron and hole gases (2DEG+2DHG), each confined to its own CrN interface. Due to the Cr $3d^3$ configuration, both electron and hole gases are based on correlated transition metal layers involving bands of $3d$ character. Transport calculations predict each subsystem will have a large thermopower, on the order of 250 $\mu$$V /K$ at room temperature. These heterostructures combine a large thermoelectric efficiency with scalable nanoscale conducting sheets; for example, operating at a temperature difference of 50K, 40 bilayers could produce a 1 V voltage with a film thickness of 100 nm.

\end{abstract}

\date{\today}

\maketitle

\section{Background}

The search for new functional materials that can be used in the fabrication of energy recovery devices based on the thermoelectric (TE) effect has been a focus of interest. \cite{disalvo,bell,snyder,dresselhaus_1} The efficiency of TE devices at a given temperature can be evaluated by using the dimensionless figure-of-merit ($ZT$), defined as $ZT$= S$^2\sigma T/\kappa$, where $S$, $T$, $\sigma$ and $\kappa$ are the Seebeck coefficient, absolute temperature, electrical conductivity, and total thermal conductivity (electron and phonon contributions), respectively. Apparently, high $ZT$ values can be achieved with concomitant high $S$ and $\sigma$ values, and low $\kappa$ values. In standard systems, $S$, $\sigma$ , and $\kappa$ are inter-related and often the optimization of one implies an undesired effect on the other, viz., large S often occurs where $\sigma$ is small.

Theoretical predictions suggesting that TE efficiency could be improved through nanostructural engineering reinvigorated interest in TE materials.\cite{dresselhaus,mahan_sofo} Using nanostructures is one possible strategy to optimize the magnitudes involved in $ZT$ by reducing the lattice thermal conductivity $\kappa$. Based on density of states (DOS) considerations, a reduced dimensionality has a positive impact in the TE efficiency due to the increased TE power given by a narrow energy distribution of the electrons participating in the transport process.\cite{dresselhaus,mahan_sofo} Decoupling thermopower and electrical conductivity is also possible, strain and nanostructuring being frequently suggested.\cite{fran_apl_materials, la2nio4_strain} 
Following the reasoning based on nano-engineered systems, surface/interface states in thin-film geometries and multilayers have drawn attention recently. The observation\cite{lao_sto_conducting} of a two-dimensional electron gas (2DEG) in LaAlO$_3$/SrTiO$_3$(LAO/STO) interfaces has stimulated excitement for the design of similar heterostructures for a variety of reasons including the enhancement of the TE power at the interface due to quantum confinement.\cite{seebeck_sto_lao,andres_nature,lao_sto_conducting,oxide_electronics_1,oxide_electronics_2,delugas}

Recently, several authors have shown that transition metal nitrides show a good TE response in addition to their excellent mechanical properties,  thermal stability, ultra-hardness, and corrosion resistance. \cite{alling,thermoel_1,camilo_adv_mat,crn_therm_1} As-grown bulk CrN exhibits a large TE power, reaching -185$\mu$$V/K$ at 420 K, an electrical resistivity that can be reduced through hole-doping in the series Cr$_{1-x}$V$_x$N, keeping a large thermopower, and a relatively low thermal conductivity due to intrinsic lattice instabilities compared to other TMN. Overall a $ZT$ of 0.04 at room temperature can be achieved, still far from the value of $ZT$ $\sim$ 1 usually required for applications. However, annealed thin films of epitaxial CrN (001) show a 250\% increase of the figure of merit at room temperature, comparable to that of PbTe or Si-Ge alloys due to their much lower electrical resistivity while high thermopower values are retained.\cite{camilo_adv_mat} These results, together with its excellent mechanical properties\cite{rivadulla_nat_mat} make CrN a promising material for high temperature TE applications.

In this paper, supported by first principles and transport properties calculations, we propose  to enhance the thermopower of nanoscale multilayers by making use of the favorable transport and mechanical properties of ionic nitrides, sandwiching CrN between (say) MgO layers along the rocksalt (111) direction. MgO is a widely used substrate material with cubic rock salt structure and a large band gap of 7.8 eV.  MgO films with atomically flat (111) surface regions, and high quality interfaces with other insulators, have been reported using layer-by-layer growth on a variety of substrates, viz. GaN(0001), 6H-SiC(0001), Al$_2$O$_3$(0001), Ag(111), and SrTiO$_3$(111).\cite{mgo_111_9, mgo_111_10, mgo_111_11, mgo_111_12, mgo_111_13, hosono}Explanation of the growth of such strongly ionic layers and quality interfaces must confront the pioneering observations of Tasker.\cite{tasker}  Based on divergent lattice sums for an ideal semi-infinite crystal with a polar surface layer, he pointed out that such surfaces must be unstable, and that substantial atomic reconstruction is one means to stabilize them. While layer-by-layer grown interfaces do contain structural imperfections (see Ref.~[\onlinecite{Millis}] for a recent discussion), interfaces with (111) orientation are reported to rival the more commonly studied (001) orientation in structural quality.  The issue of course is the establishment of an internal electric field, for which there have been model\cite{mgo_111_2,mgo_111_3,mgo_111_9,Noguera2008} and first principles\cite{Wander2001,mgo_111} studies.  We return to these questions in the Discussion section. 

Considering that there are several reports of preparation of atomically 
flat MgO(111) surface areas, 
MgO is promising as a substrate to explore the possibility of polar interface engineering using 
mononitrides with rocksalt structure.
Bulk CrN is a paramagnetic semiconductor with cubic rocksalt structure at room temperature that 
transforms to orthorhombic and antiferromagnetic below 285 K. The magnetic configuration consists 
of double ferromagnetic (FM) layers stacked antiferromagnetically along the [110] 
direction (AFM2).\cite{crn_films_asbotana, crn_magnetism_bulk, prb_crn_ours} 

Recently, it has been theoretically predicted\cite{ours} that ScN/MgO(111) multilayers designed as a superlattice constitute a new type of alternating metal-insulator multilayers with electron and hole 2DGs confined to the two opposing interfaces of the structure (beyond a critical thickness of ScN necessary for inducing interfacial metallicity). The ionic picture of CrN/MgO(111) multilayers is similar to the ScN case, leading to comparable polar distortions that give rise to alternating electron+hole conducting gases in subsequent interfaces but with an enhanced thermopower coming from the narrower Cr $3d$ bands around the Fermi level, which should also provide a large conductivity, particularly for the hole gas.

\section{Computational Details}
Electronic structure calculations were performed within density functional theory\cite{dft,dft_2} using the all-electron, full potential code WIEN2k\cite{wien2k} based on the augmented plane wave plus local orbitals (APW+lo) basis set.\cite{sjo} For the structural relaxations we have used the Wu-Cohen version of the generalized gradient approximation (WC-GGA)\cite{wu_cohen} that gives better lattice parameters for MgO than PBE.\cite{tran_wc}

To deal with strong correlation effects, we apply the LDA+U scheme\cite{sic} that incorporates an on-site Coulomb repulsion U and Hund's rule coupling strength J$_H$ for the Cr-3d states. The LDA+U scheme improves over GGA or LDA in the study of systems containing correlated electrons by introducing the on-site Coulomb repulsion U applied to localized electrons. The value U= 4 eV was used for CrN, since it has been shown to give a reliable  bulk spectrum\cite{herd,prb_crn_ours} based on the agreement with photoemission experiments and transport properties, the comparison between band structure parameters and optical data, and a comparison with parameter-free functionals. The on-site Hund's exchange parameter was set as J$_H$= 0.7 eV. 
All calculations were well converged with respect to all the parameters used. In particular, we used $R_{mt}K_{max}$= 7.0, and muffin-tin radii were 1.95 a.u. for Cr, 1.73 a.u. for N, 1.79 a.u. for Mg and 1.69 a.u. for O. The calculations used a 43$\times$43$\times$5 $k$-mesh for the integrations over the Brillouin zone.
We have modeled MgO/CrN(111) multilayers (1$\times$1 in plane) with thickness of the CrN block being varied from two to seven CrN layers. A barrier of 4-5 MgO layers (about 1 nm thick) between CrN blocks has been used for all the calculations checking that it is enough to guarantee the lack of interaction between CrN blocks. AFM order along (111) was set in the CrN blocks. Note that the imposed AFM ordering along (111) is not the bulk CrN AFM2 ordering since it is not well adapted to the confinement constraint of these thin layers.\cite{crn_films_asbotana,crn_magnetism_bulk} 

The in-plane lattice parameters are constrained to those of MgO (fixed to the cubic lattice constant 4.23 \AA, obtained optimizing the cell volume within WC-GGA). CrN grown on top of MgO is under tensile strain since its lattice parameter is approximately 2\% smaller than that of MgO. 

The transport properties were calculated using a semiclassical solution based on Bloch-Boltzmann transport theory within the constant scattering time approximation by means of the BoltZTraP code,\cite{boltztrap} which uses the energy eigenvalues calculated by the WIEN2K code. The constant scattering time approximation assumes that the energy dependence of the scattering time at a given temperature and doping level is negligible on the scale of k$_B$T. In this case, denser k-meshes are required, in our case up to 118$\times$118$\times$14 to reach convergence.

\section{results}

\begin{figure}
\center
\includegraphics[width=0.8\columnwidth,draft=false]{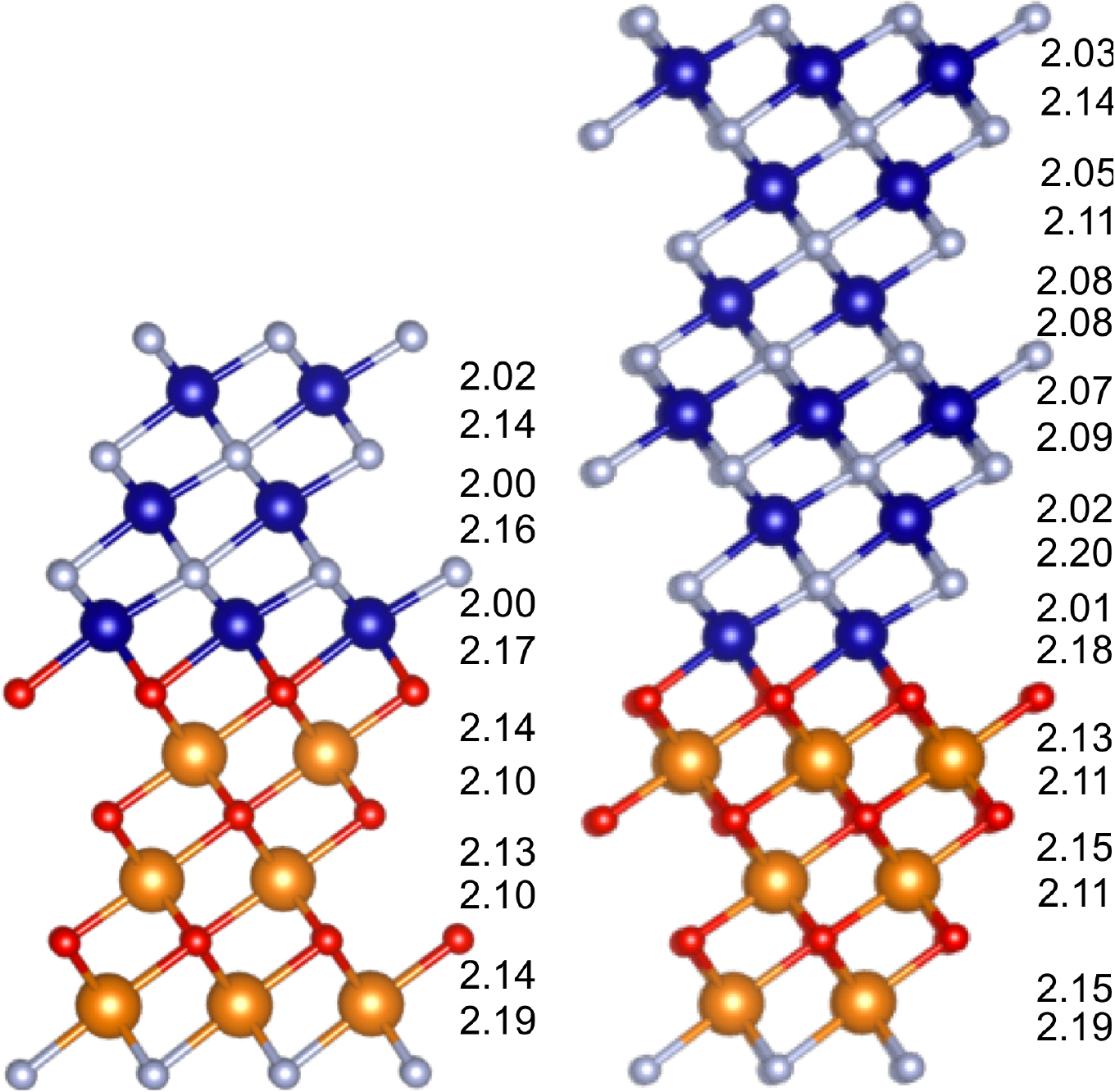}
\caption{Relaxed structure of MgO/CrN (111) multilayers 3- and 6- CrN layers thick with Cr atoms in blue, N atoms in gray, Mg atoms in orange and O atoms in red. The average distances between CrN layers along the z axis (in \AA)
are shown.}\label{sup1}
\end{figure}

\begin{figure}
\center
\includegraphics[width=\columnwidth,draft=false]{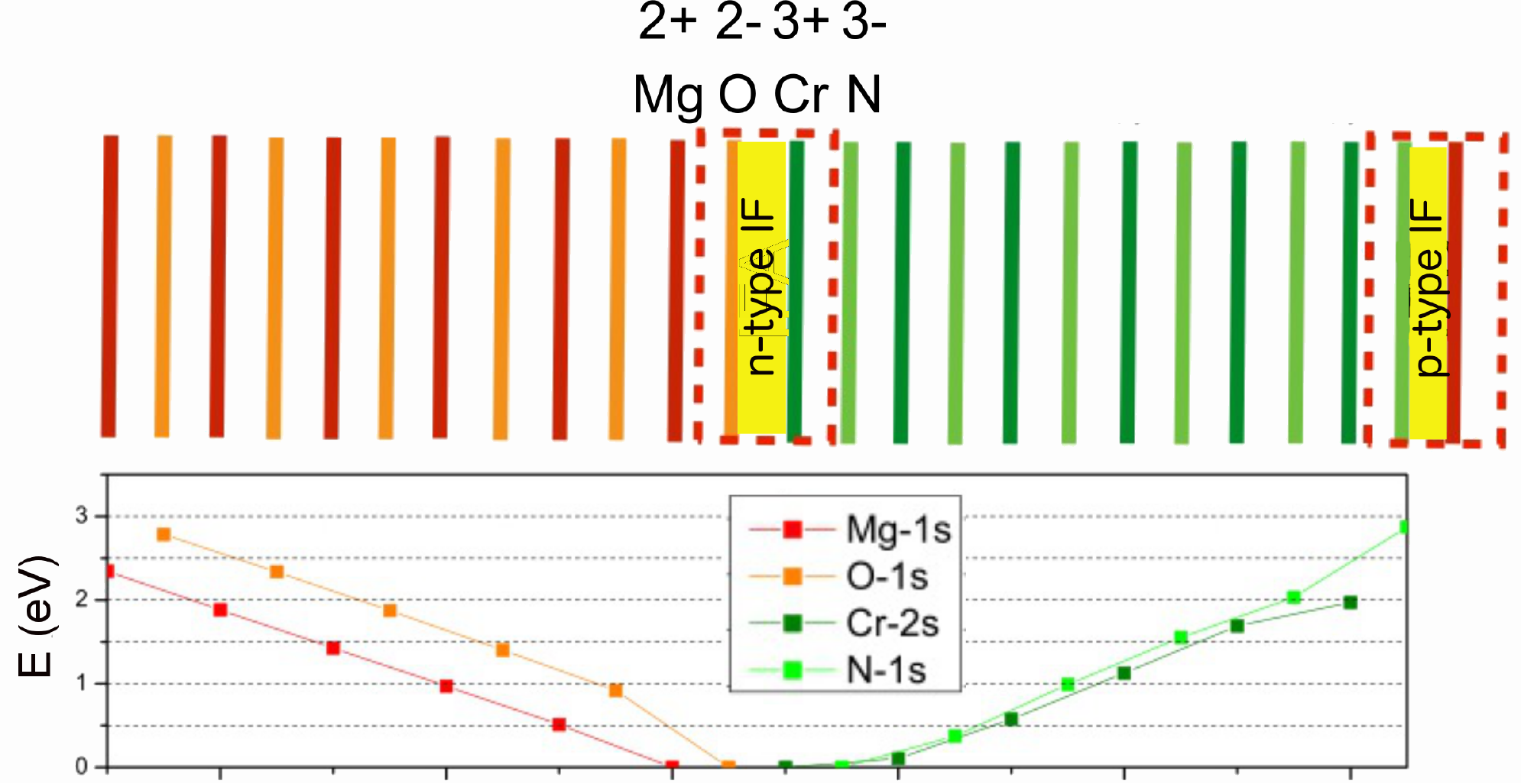}\\
\caption{ (Top panel) Scheme showing the formal charge of the layers in a superlattice 6-CrN layers thick: 3+ for Cr (dark green), 3- for N (light green), 2+ for Mg (red), and 2- for O (orange). (Bottom panel) Potential gradient shown as the shift in the Mg/N/O-1s and Cr-2s core states layer-by-layer for a CrN/MgO(111) multilayer 6 CrN-layers thick. The zero is set at IFA (n-type). Cr-2s (dark green), N-1s (light green), Mg-1s (orange), and O-1s (red).}\label{fig2}
\end{figure}

\textit{Polar distortions.} The (111) orientation of a rocksalt structure is polar and a crystal cut along (111) presents alternating layers of metal and oxygen/nitrogen ions forming triangular sublattices in the ab plane.  Cr and N planes have alternating formal charge 3+ and 3-, with 2+ and 2- for MgO (see Figs. \ref{sup1}, \ref{fig2}). Hence, the heterostructure contains two charge imbalanced IFs: one $n-$type (IFA, 2-/3+) and one $p-$type (IFB, 3-/2+). 

For all the multilayers we performed calculations with fully relaxed atomic positions also optimizing the value of the $c$-lattice parameter (off-plane), i.e. allowing atomic displacements along the $c$-axis and thus relaxing the inter-plane distances for the structures with different number of CrN layers. The optimized values of the $c$-lattice parameter result in a slight decrease from the value obtained by constructing the multilayers from the MgO bulk unit cell, of about 1\% to  2\% for multilayers two to seven CrN layers thick. This is expected, as the in-plane lattice parameter inside the CrN blocks is constrained to that of MgO, which is larger (4.23 \AA~ versus 4.13 \AA~ for the cubic phase of CrN). 

One of the main features in the calculated relaxed geometry is the presence of a considerable polar distortion along the (111) direction with the atomic displacements being exclusively along the $c$-axis. On the CrN side they result in one shorter (2.01-2.07 \AA) and one longer (2.14-2.18 \AA) Cr-N or Cr-O bond length as can be seen in Fig. \ref{sup1}. The distances are balanced in the inner layers for the thicker CrN blocks somehow recovering the nearest neighbor distance in bulk CrN of 2.07 \AA. The distortions are also present on the MgO side with Mg-O, Mg-N distances from 2.11 to 2.19 \AA.

The polarity in MgO/CrN (111) multilayers gives rise to a large electrostatic potential offset between the two interfaces, proportional to the thickness of the CrN block. This internal potential can be tracked easily through the core levels, as shown in Fig. \ref{fig2}. A potential difference of 2 to 3 eV arises across the CrN and MgO slabs, increasing with increasing thickness of the CrN block. The core level shifts are about 1 eV/layer in the thinner multilayer and are reduced as the CrN block becomes thicker. The fact that the inner layers of the thicker CrN slabs show no polar displacements (as discussed above) indicates that the internal electric field has been mitigated. 

\begin{figure*}
\center
\includegraphics[width=2\columnwidth,draft=false]{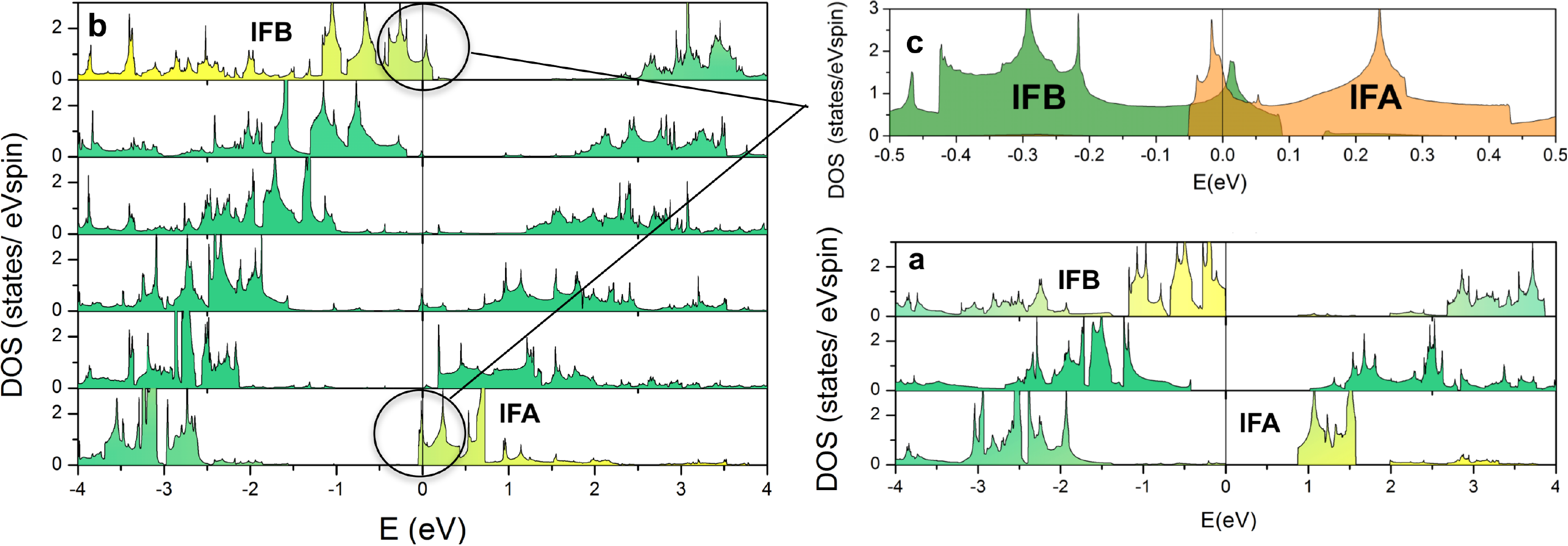}\\
\caption{Layer resolved DOS for Cr atoms across the CrN slab from IFB to IFA for CrN/MgO(111) multilayers three (a) and six (b) CrN-layers-thick showing the Cr-d states. The plots show the DOS for the majority spin. The shift  in the Cr-t$_{2g}$ peak as moving from IFB to IFA can be observed. The (c) panel shows a zoomed density of states around the Fermi level for a multilayer 6-CrN layers thick.}\label{fig3}
\end{figure*}

\textit{Electronic structure.} The band gap of the system decreases with increasing thickness of the CrN film before an insulator-to-metal transition takes place at 4 CrN layers, as a consequence of the potential gradient due to interface polarity. The transition between insulating and conducting states is hence a sharp function of the thickness, as was predicted and confirmed experimentally in LAO/STO heterostructures.\cite{science_thiel,pentcheva_prl,hole_mobility}

The density of states (DOS) for the case of a system 3-CrN layers thick (insulating state) is depicted in Fig. \ref{fig3}(a). The electronic structure can be understood from a simple ionic picture: octahedral Cr$^{3+}$:d$^3$ with spin 3/2, and the $e_g$ states at higher energy. The gap lies between occupied Cr(IFB)-t$_{2g}$ states and unoccupied Cr(IFA)-e$_g$ states. From the layer-by-layer DOS for the 6-CrN case shown in Fig. \ref{fig3}(b), it is evident that the interfacial $t_{2g}$ and $e_g$ states are affected differently compared to interior layers. Thus the gap widening is better viewed as interface differentiation of the crystal field split Cr $d$ states. The layer-by-layer shift from IFA to IFB (of 0.7 eV per Cr layer) arises from  the internal electric field. Upon increasing the number of CrN layers, the gap  between bands at IFA and IFB closes,  with the insulator to metal transition occurring  at a critical thickness of 4 CrN layers. When metallicity arises, the potential gradient is reduced significantly with respect to the thinner CrN layers, due to shift of electrons across the CrN slab.  Band overlap of filled Cr $t_{2g}$ bands at IFB with empty Cr $e_g$ bands at IFA results in 2D bilayer conducting (hole+electron) gases, each confined to its own interface. 

\begin{figure}
\center
\includegraphics[width=\columnwidth,draft=false]{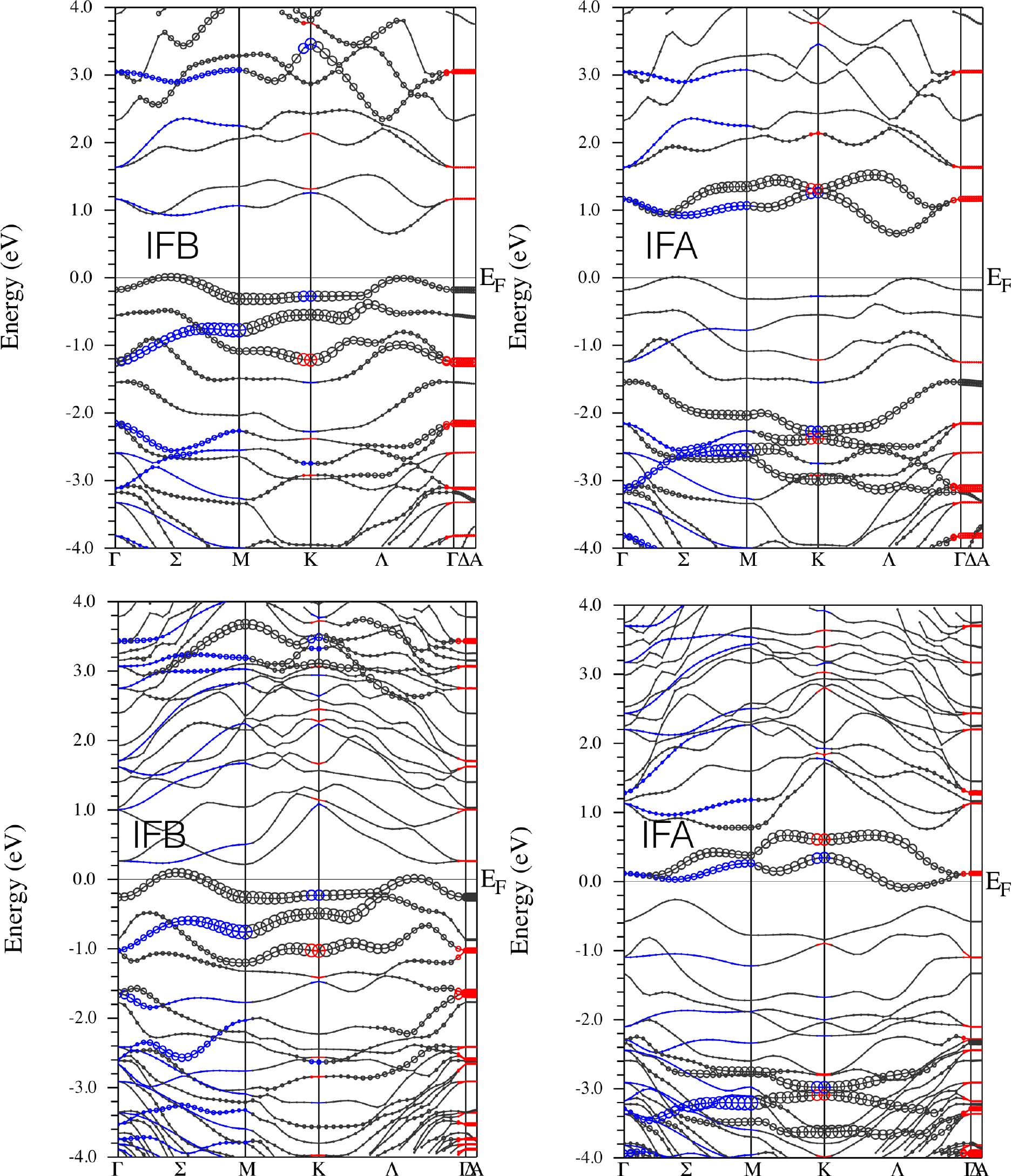}\\
\caption{Band structure with band character plot for MgO/CrN (111) multilayers. Cr-$e_g$ at IFA(right) and Cr-$t_{2g}$ levels at IFB (left) for CrN/MgO(111) multilayers three (top panel) and six (bottom panel) CrN-layers-thick.}\label{bs}
\end{figure}

Figure \ref{bs} shows the band structures with band character plot that complement Fig.\ref{fig3}. The bands crossing the Fermi level are indeed Cr-$t_{2g}$ at  IFB and Cr-$e_g$ at the IFA.  At the interface, the octahedral environment of the Cr atoms is clearly distorted as can be observed in Fig. \ref{sup1}. Due to the trigonal distortion there is a splitting of the t$_{2g}$ levels into a lower lying e$_g$' doublet and a higher lying a$_{1g}$ singlet. The a$_{1g}$ band is flatter in character and gives rise to the 1D-like DOS. The e$_g$ levels are unaffected by the trigonal distortion.

At first glance the basic electronic structure seems similar to that of ScN/MgO multilayers. However, here the states involved at both interfaces are the much narrower Cr-$3d$ bands, and this difference  produces substantial enhancement of the TE response.  Unlike other systems studied previously, the hole gas is formed by Cr-$3d$ bands instead of N-$p$ bands or O-$p$ bands. The low (or zero) mobility of the holes at the $p-$type interface of LAO/STO multilayers leads to a negligible conductivity for that interface.\cite{hole_mobility} While that situation is improved in ScN/MgO (111) interfaces,\cite{ours}  CrN/MgO is qualitatively different due to the sole involvement of $d$-electron bands.

\begin{figure}
\center
\includegraphics[width=\columnwidth,draft=false]{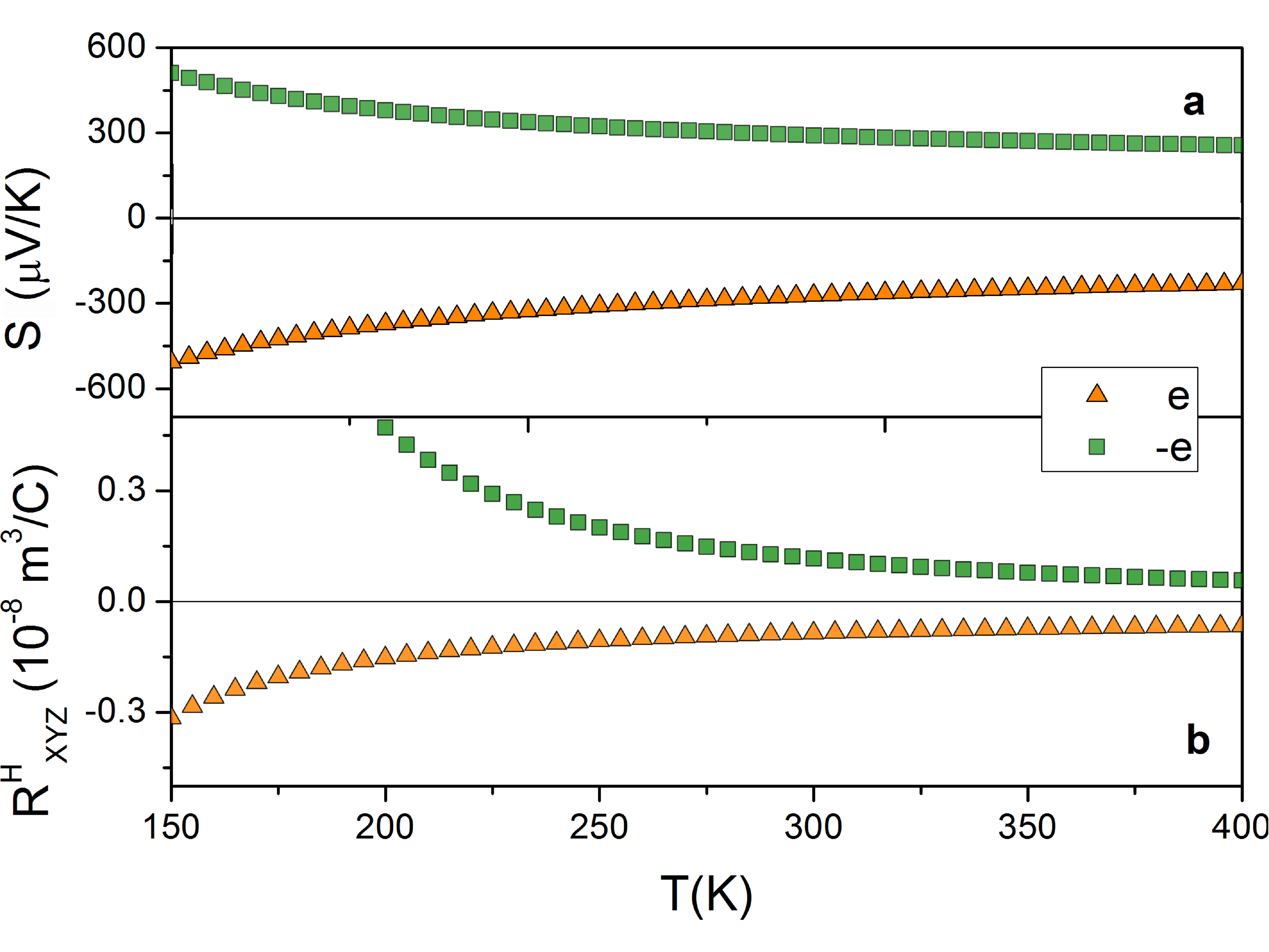}\\
\caption{ Temperature dependence of the in-plane thermopower (a) and Hall coefficient (b) for MgO/CrN multilayers six CrN layers thick for electrons and holes. The contributions to the net $S_{xx}(T)$ are discussed in the text. The calculations were done by summing over only the valence Fermi surfaces for the hole properties (green), only over the conduction Fermi surfaces for the electron properties (orange). At lower temperature S(T) turns over and goes linearly to zero.}\label{fig4}
\end{figure}

\textit{Transport properties.} To check the TE response in this type of two-carrier 2DG multilayered system, we have carried out transport property calculations based on Bloch-Boltzmann transport theory in the constant relaxation time approximation. Within this approximation, the TE power is independent of the scattering time ($\tau$). 
 When a conventional two-carrier electron-hole system develops as in simple semimetals,  hole and electron contributions to the Seebeck coefficient  compensate, often resulting in a small, weakly temperature dependent, and uninteresting thermopower. However, for these physically separated electron gases, it is appropriate and necessary to compute separately the contributions from each 2DG, i.e. from the electron pockets, then the hole pockets. The thermopower calculated from the  semi-classical Bloch-Boltzmann expressions,\cite{Allen1988} which involve  only Fermi surface quantities, is shown in Fig. \ref{fig4}(a). Values of the TE power for electrons and holes are, accidentally, virtually identical in magnitude (both gases come from $3d$ electron bands of different symmetry), with the independent thermopower at each interface reaching very high values of about 250 $\mu$V/K above 300 K where it becomes T independent. The DOS of electron and hole bands separately indicates that the enhanced thermopower comes from an almost 1D-like DOS arising at both  interfaces (see Fig. \ref{fig3}(c)). 

If one can complement this large thermopower with a large conductivity and a reduced thermal conductivity, this multilayered system can exhibit a TE performance comparable to that of the materials used in commercial TE devices. In our calculations within the constant relaxation time approximation,  $\sigma$/$\tau$ and $\kappa$/$\tau$ are obtained. Hence, the TE figure of merit $ZT$ is independent of the scattering time as the dependence on $\tau$ cancels out in $\sigma$/$\kappa$.  It should be noted that this estimate contains the electronic contribution only since no phonon terms are considered for the thermal conductivity. Thus, it can be considered as a theoretical upper limit of the figure of merit. The estimated $ZT$ at room temperature for each 2DG separately is $\sim$ 0.9 in the threshold required for applications. A more realistic estimate of the figure of merit for this system can be made using the TE power we have calculated together with the experimental data for $\sigma$ and $\kappa$ taken from Ref.\onlinecite{camilo_adv_mat} giving rise to a value of 0.64 at room temperature.

The Hall coefficient is also $\tau$ independent within the constant relaxation time approximation. The calculated values (see Fig. \ref{fig4}(b)) for electron and hole-like bands give $R^H_h$ = 0.06$\times$10$^{-8}$ $m^3/C$, $R^H_e$ = -0.06$\times$10$^{-8}$ $m^3/C$, both varying slowly with T above 300 K. Using a 2DG thickness of 3 \AA, these values correspond in a standard parabolic single band interpretation to an areal carrier density $n_e$=$n_h$ of 3.1$\times$10$^{14}$ carriers/cm$^2$. This value is similar in magnitude to the electronic density (from the Hall number) seen at the $n-$type LAO/STO interface at a large number of LAO overlayers.\cite{caviglia,pentcheva_hole_electron}  

\section{Discussion and Summary}
\subsection{Some technical considerations}
The parallel superlattice 2D(E+H)Gs of the type discussed here corresponds to alternately layered $n-$ and $p-$type 2DGs, each with its own Hall voltage and Seebeck coefficient. This flexibility, plus the ability to vary the number of CrN layers in the heterostructure, provide a high degree of tunability that can be further controlled in a gate-voltage device. The large TE response that can be provided by the two carrier 2D gases independently, makes this multilayered system promising as a TE device based on a nanoscale architecture. With S(250K) = 250 $\mu$V/K operating across a 50 K temperature difference, only 40 bilayers (as little as 100 nm) are necessary to create a 1 V voltage difference. 

\begin{figure}
\center
\includegraphics[width=0.8\columnwidth,draft=false]{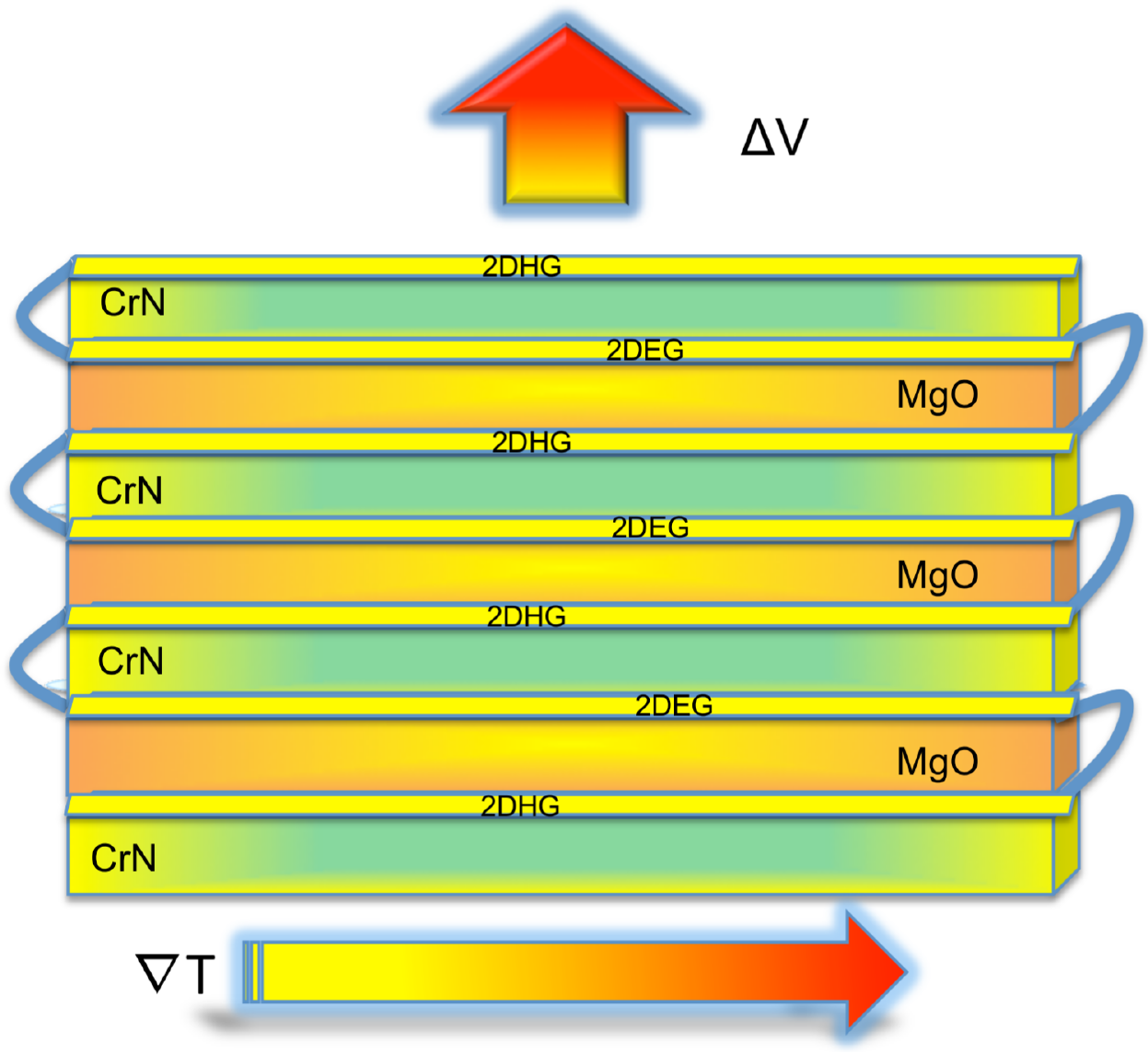}\\
\caption{Sketch of the thermoelectric module proposed using a nitride-oxide platform consisting of parallel two-dimensional electron and hole gases (2DEG and 2DHG, respectively) created by interfacial confinement in a multilayered structure. The thermal gradient flows as shown, perpendicular to the voltage. The $n$ and $p-$type sectors are connected in series (with the wires represented pictorially), as usual in the type of thermoelectric modules we describe in the text.}\label{fig1}
\end{figure}

TE devices contain many TE couples consisting of $n-$type (electron carriers) and $p-$type (hole carriers) TE elements wired electrically in series and thermally in parallel. In the multilayered system proposed here, with the two gases showing a large TE response, one could envisage a TE module constructed as shown in Fig. \ref{fig1}, where each 2DG acts as the $n-$ or $p-$type element of the TE couple. This scheme presents several intrinsic advantages for a nanoscale TE device. Such a device is intrinsically nanostructured (since the internal voltage scales with the nanostructuring of the device, and the $n-$ and $p-$type parts can be isolated with just a 1-2 nm MgO barrier). The design proposed here would have very good  compatibility between the $n-$type and $p-$type materials (the match of electrical and thermal properties of p- and n-materials required for an optimum efficiency of the TE device is often a challenging problem). Mechanical stability is another strong point of this nitride-oxide system and scalability and nanoscale confinement of the 2DG can be controlled using atomic force microscopy lithography, enabling fabrication of nanoelectronic devices operating at the interface.\cite{oxide_electronics_1,oxide_electronics_2}  

\subsection{Growth  of polar materials}
Synthesis of the multilayers we have proposed requires growth of charged layers a few nanometers thick. Such growth at first appears daunting. Tasker established the widely known result that an ideal ionic surface of a semi-infinite polar crystal is unstable, based on the divergence of lattice sums. Yet, as mentioned in the Introduction, several groups have established growth of charged (111) layers with interface quality that rivals that of neutral layers, based on transmission electron microscopy and measured properties. Tasker proposed surface reconstruction (of the atomic geometry and hence ionicity) as one means of stabilizing ionic surfaces.\cite{tasker} The questions posed by observation of flat polar surfaces began to become more specific when Wander {\it et al.} made a theoretical investigation of ZnO(0001) surfaces.\cite{Wander2001}  There have since been several suggestions of ways to achieve flat polar nanostructures.

First, the semi-infinite geometry can be avoided. Growing (ultra)thin layers of the starting ``substrate'' has met with success. Matsuzaki {\it et al.} observed layer-by-layer growth of MgO(111) thin films on non-polar NiO buffer layers up to 10 MgO layers.\cite{mgo_111_12,mgo_111_13,hosono} Only for thicker layers step and terrace structures appear, presumably to counteract the growing dipole. Noguera and Goniakowski had performed model studies of such ultrathin films, concluding\cite{mgo_111_2,mgo_111_3,Noguera2008} that both surface (interface) charges and substantial lattice distortions can play a role in compensating the electric dipole that increases with thickness. Surface or interface charges in MgO(111)/Ag(111) were also identified by Arita {\it et al.} as a mechanism of mitigating the increasing dipole in ultrathin MgO(111) films.\cite{mgo_111} It had earlier been reported that unreconstructed wurtzite ZnO(0001) surfaces of thin films were assisted in stabilization by surface charges.\cite{Wander2001}
Surface composition and roughening during growth was identified by Benedetti {\it et al.} as a means to compensate the growing dipole in MgO(111) films.\cite{Benedetti2011}  Different growth regimes can be accessed by control of oxygen pressure, growth temperature, and availability of water vapor in the growth environment.

A key feature of the system we propose is that only alternating {\it ultrathin layers} of MgO and CrN are grown, and that the MgO(111) ``substrate'' can also be an ultrathin layer grown on a convenient nonpolar substrate. For the conducting interfaces that we envision, the metallic interface layers will provide a polarizability that will help to mitigate internal electric fields not only during growth but in the completed system as well. Finally, treatment of the uppermost surface may enhance stability of the atomic and electronic structure. The question of whether cations and anions can be grown sufficiently ordered (that is, with only minor intermixing) remains one for experiments to resolve.

\subsection{Summary}

To summarize, here we show an enhancement in the TE response in rocksalt nitride (CrN)/ oxide (MgO) superlattices grown along the polar (111) direction containing one $n-$type and one $p-$type interface. This multilayered system has four distinctive features compared
to the more highly studied multilayers. First, the (111) growth
orientation, which very recently is being shown to provide high
quality interfaces. Secondly, the $d^3$ configuration of the Cr
ion dictates that both 2DGs have $d$ electron character:
holes lie in the $t_{2g}$ bands, with the electrons being promoted
to the $e_g$ bands. Thirdly, both types of carriers lie in a
transition metal {\it nitride} rather than oxide, which should
provide high thermopower (on the order of 250 $\mu$V/K at room temperature) and a conductivity larger than common in oxides, particularly
for the holes, which often localize at oxide interfaces.
Finally, in a point that we have not emphasized, the carriers
are fully spin polarized, with the magnetism being determined not 
by interface physics but by bulk properties, which will make it
much more robust. It also raises the possibility of control of
properties by a magnetic field, which is a topic for future study. Based on this configuration we propose a new type of nanoscale TE device consisting on ultrathin nitride layers carrying electrons and holes at alternating interfaces. Such a device would provide significant advantages in terms of improvement of standard losses and mechanical resistance issue as well as a vast potential for scalability.

\section{acknowledgments}

V. P. thanks MINECO for project MAT2013-44673-R, the Xunta de Galicia through project EM2013/037, and Spanish Government for financial support through the Ram\'on y Cajal
Program (RyC-2011-09024). W.E.P. acknowledges many conversations with R. Pentcheva on the theory and
phenomenology of oxide interfaces and multilayers. 
A.S.B. 
 and  W.E.P were supported by Department of Energy Grant No. DE-FG02-04ER46111. This research used resources of the National Energy Research
Scientific Computing Center, a DOE Office of Science User Facility
supported by the Office of Science of the U.S. Department of Energy
under Contract No. DE-AC02-05CH11231.

\end{document}